A manuscript entitled

# CONVECTIVE HEAT TRANSFER ENHANCEMENT OF NON-CONDENSING VERTICAL UPWARD SLUG FLOW USING CAPILLARY-ASSISTED PHASE SEPARATION

Milad Darzi

Chanwoo Park*

Department of Mechanical and Aerospace Engineering

University of Missouri, Columbia, MO 65211, USA

* Please address all correspondence to:

Chanwoo Park

Associate Professor

Department of Mechanical and Aerospace Engineering

University of Missouri, Columbia

E2402 Lafferre Hall

Columbia, MO 65211

Phone: +1-573-882-6087

Fax: +1-573-884-5090

E-mail: parkchanw@missouri.edu

**Abstract**

This paper discusses the convective heat transfer enhancement of non-condensing slug flows by using phase separation for local liquid circulations through a porous-tube insert. Numerical simulations were carried out to study the air-water slug flow and heat transfer without phase change (boiling and condensation) in a vertical upward flow. The governing equations of two-phase flow and heat transfer were numerically solved using the Volume of Fluid-Continuum Surface Force (VOF-CSF) method in a two-dimensional axisymmetric computational domain. The rear end (exit) of the porous tube was open to allow discharge of fluid, but two different designs of the frontal end were considered; a porous tube with a closed front that prevents the bubbles from entering the porous tube and keeps the bubbles in the annular gap; a porous tube with an open front allows the bubbles to flow through the porous tube. Simulation results show that a counter flow of the bulk liquid inside and outside the porous tube creates internal liquid circulations in both the axial and radial directions, resulting in an increased liquid velocity near the tube wall, thus an enhanced convective heat transfer. It is also found that there exists an optimum diameter of the porous-tube-insert (or annular gap dimension between the tube wall and porous tube) for convective heat transfer enhancement. The maximum Nusselt number of the porous-tube-insert cases used in this study was about five times higher than that of the liquid-only flow in a bare tube.

**NOMENCLATURE**

| | |
|---|---|
| $A$ | cross-sectional area [m$^2$] |
| C | Courant number, C=$u\Delta t/\Delta x$ [-] |
| $c_p$ | specific heat capacity [J/kg-K] |
| $D$ | tube dimeter [m] |
| **$F$** | force per length [N/m] |
| $g$ | gravity [m/s$^2$] |
| $L$ | length [m] |
| $k$ | thermal conductivity [W/m-K] |
| Nu | Nusselt number [-] |
| $p$ | pressure [Pa] |
| Pr | Prandtl number [-] |
| $q''$ | heat flux [W/m$^2$] |
| $r$ | radial coordinate, radius of porous tube [m] |
| $R$ | tube radius [m] |
| Re | Reynolds number [-] |
| $t$ | time [s] |
| $T$ | temperature [°C] |
| ***u*** | velocity [m/s] |
| $x$ | vapor quality [-] |
| $X$ | Martinelli parameter [-] |
| $z$ | axial coordinate [m] |

**Greek Letters**

| | |
|---|---|
| $\alpha$ | phase fraction in a computational cell [-] |
| $\varepsilon$ | void fraction [-], $\varepsilon=V_G/(V_G+V_L)$ or $\varepsilon=A_G/(A_G+A_L)$ |
| $\mu$ | dynamic viscosity [Pa/s] |
| $\upsilon$ | kinematic viscosity [m²/s] |
| $\rho$ | density [kg/m³] |
| $\delta$ | liquid film thickness [m] |
| $\sigma$ | surface tension [N/m²] |
| $\Delta$ | annular gap dimension [mm] |

**Subscripts**

| | |
|---|---|
| $1\varphi$ | single-phase |
| $2\varphi$ | two-phase |
| $b$ | bulk, bubble |
| $G$ | gas |
| $L$ | liquid |
| $s$ | liquid slug |
| $su$ | slug unit |
| $w$ | wall |

**Symbols**

| | |
|---|---|
| $\overline{\phantom{x}}$ | time-averaged |

## 1. Introduction

Heat transfer enhancement has been always a constant endeavor of scientists and engineers aiming to reduce energy consumptions for maximum efficiency. For liquid-phase flows, the addition of a secondary different phase such as solid particles or gas bubbles has been shown to be very effective in enhancing the convective heat transfer due to the increased flow mixing and boundary layer disturbance of the multi-phase flows.

Hassanipour and Lage [1] used a liquid flow laden with phase-change solid particles in a mini-channel for a moderate (20% increase) heat transfer enhancement, as compared to the liquid-only (without particles) flow. The enhancement was attributed to the combination of a flow mixing induced by the solid particles and a phase-change effect. Kiaee and Lage [2] numerically analyzed a liquid flow loaded with a spherical solid particle in a channel to study the role of the particle sweeping the wall boundary layers on the convective heat transfer and reported an increase of 12% in the average Nusselt number. Kiaee et al. [3], using a numerical simulation showed that the presence of solid particles in a liquid flow led to enhancements by 27% and 10% in the average Nusselt number and average heat flux, respectively.

For gas-liquid flows without phase change (i.e. non-boiling/non-condensing), a slug flow can significantly increase the convective heat transfer compared to the conventional liquid flow. A gas-liquid flow in a channel may exist in various flow regimes depending on the velocities of the phases and flow orientation. The slug flow (aka bubble-train flow, Taylor flow, and intermittent flow) is an important flow regime that is characterized by the appearance of bullet-shape bubbles, separated by liquid slugs. Prothero and Burton [4] from an experimental study found that the heat transfer of air–water slug flows in capillaries were twice of that of a liquid-only flow. Oliver and Hoon [5] reported the two and half times heat transfer enhancement for a gas-liquid slug flow in a channel of diameter 6.4 mm compared to that of a liquid-only flow.

Lakehal et al. [6] performed numerical simulations of air-water flow in a miniature tube of diameter 1 mm and showed that the air-water slug flow increases the convection heat transfer by four times more than that of a liquid-only flow. Gupta et al. [7] numerically simulated air-water slug flows in a channel of diameter 0.5 mm and found that the Nusselt numbers of the slug flow were two and a half times higher than those of liquid-only flow for both constant wall heat flux and constant wall temperature conditions. Mehdizadeh et al. [8] using the VOF method simulated air-water slug flows in a mini-channel of 1.5 mm diameter under a constant wall heat flux and found that the Nusselt number of the slug flow can be 610% higher than that of single-phase flow. They identified two main reasons contributing to the enhanced heat transfer of the slug flow [9]; first, the secondary flow recirculation in the liquid slugs to promote the radial heat transfer; second, higher local velocity of the liquid slugs due to the reduced liquid fraction.

It has been shown in literature that, two immiscible liquids enhances the convective heat transfer in two-phase flows. Gat et al. [10] in an experimental study showed that an induced liquid-liquid phase separation can enhance the convective heat transfer up to 130% as compared to that of a single-phase flow. Xing et al. [11] performed an experimental study in a microchannel and showed that when heated above a critical temperature, the mixture separated into distinct immiscible liquid phases resulting in an enhancement by 250% in the convective heat transfer coefficient as compared to that of a single liquid-phase flow. Ullmann [12] conducted an experimental study of laminar flows in a small tube diameter of 2 mm and found that the induced liquid-liquid phase separation can substantially enhance the convective heat transfer up to 150%. The heat transfer enhancement was attributed to the increased fluid mixing and the lateral movement of the separating droplets.

In condensing vapor-liquid flows, the heat transfer inherently degrades toward the condenser outlet where more condensate is accumulated, and thus more condenser surface is covered by a thick condensate layer, resulting in a large thermal resistance. To mitigate this problem, Chen et al. [13] utilized a gas-liquid phase separation using a porous tube to promote a thin-film condensation in the condenser. The phase separation is achieved by capillary force at the liquid-vapor interfaces (aka meniscus) formed in the pores of the porous tube. The porous tube blocks the gas bubbles from entering the porous tube, but allows the liquid to freely flow through the porous tube. Consequently, the confined bubbles in the annular gap between the condenser wall and porous tube, which is elongated along the condenser wall, promotes the thin film condensation in the condenser. Chen et al. [13] observed such a phase separation in an adiabatic (no heat transfer) horizontal air-water flow experiment. Also, Chen et al. [14] from the aforementioned research group using a numerical simulation, investigated the hydrodynamics of an adiabatic, vertical upflow, air-water slug flow with the presence of a porous-tube-insert. The numerical results showed a favorable phase separation and ultra-thin liquid film near the tube wall leading to condensation enhancement. Moreover, secondary liquid circulations in the radial direction were observed over the entire tube length.

In this paper, a convective heat transfer enhancement using non-condensing gas-liquid flows and a capillary-assisted phase separation was numerically investigated. Air-water slug flows in a vertical upward flow arrangement was used to study the hydrodynamics of the slug flows and fundamental mechanisms of the convective heat transfer enhancement by the phase separation using a porous-tube-insert. Two designs of the porous tube were considered: a closed frontal end to prevent bubbles from entering the porous tube; and an open front end to allow bubbles to enter and flow through the porous tube. A set of parametric simulations was systematically carried out for an optimization of the annular gap dimension (or porous tube diameter) for the maximum heat transfer.

## 2. Computational model

### 2.1 Governing equations

The Volume-of-Fluid (VOF) method was used to simulate a transient gas-liquid two-phase flow [15]. In this modeling, each phase is treated as an incompressible immiscible fluid and the phase distribution is determined using phase fraction ($\alpha$) that is a scalar variable for the phase fraction of one phase in a computational cell. Both fluids are assumed to be Newtonian. No phase change was considered. The governing equations for mass, momentum, and energy conservation are given by

$$\frac{\partial}{\partial t}(\rho) + \nabla \cdot (\rho \mathbf{u}) = 0 , \tag{1}$$

$$\frac{\partial}{\partial t}(\rho \mathbf{u}) + \nabla \cdot (\rho \mathbf{u}\mathbf{u}) = -\nabla p + \nabla \cdot (\mu \nabla \mathbf{u}) + \rho \mathbf{g} + \mathbf{F}_\sigma , \tag{2}$$

$$\frac{\partial}{\partial t}(\rho c_p T) + \nabla \cdot (\mathbf{u} \rho c_p T) = \nabla \cdot (k \nabla T), \tag{3}$$

In addition, the phase fraction ($\alpha$) within a cell and the motion of interface is computed using an additional transport equation:

$$\frac{\partial}{\partial t}(\alpha) + \nabla \cdot (\mathbf{u}\alpha) + \nabla \cdot [\mathbf{u}_r \alpha (1-\alpha)] = 0 . \tag{4}$$

In the above equations, $\mathbf{u}$ denotes the velocity vector, $p$ the pressure, and $T$ the temperature. The bulk properties such as density $\rho$, viscosity $\mu$, thermal conductivity $k$, and heat capacity $c_p$ are calculated using the average property values of the two phases, weighted by their respective phase fractions $\alpha$ [i.e. $\rho = \alpha_L \rho_L + (1-\alpha_L)\rho_G$]. The surface tension ($\mathbf{F}_\sigma$) is approximated as a body force in the momentum equation [Eq. (2)] and is computed using the continuum surface force (CSF) model in the vicinity of the interface ($0 < \alpha < 1$) as proposed by Brackbill et al. [16] and is given by

$$\mathbf{F}_\sigma = \sigma \kappa \nabla \alpha , \tag{5}$$

where $\sigma$ is surface tension and $\kappa$ is interface curvature defined by $\kappa = \nabla \cdot (\nabla \alpha / |\nabla \alpha|)$. To mitigate the numerical smearing of the interface, an extra artificial compression term is incorporated in the phase fraction equation [third term in Eq. (4)] to sharpen the interface [17]. The compression term includes a compression velocity $\mathbf{u}_r$ that is available only in the normal direction to the interface. The $\mathbf{u}_r$ is based on the velocity magnitude in the transition region of the thin interface and is determined with a maximum velocity ($u_{max}$) and an adjustable coefficient ($K_c$) defining the extent of the compression and a normal vector of the interface and is given by

$$\mathbf{u}_r = K_c U_{max} \frac{\nabla \alpha_k}{|\nabla \alpha_k|} . \tag{6}$$

where the adjustable coefficient ($K_c$) can vary between zero and four. In this study, $K_c$ is set to unity as suggested in literature [18]. The VOF model including the additional artificial compression term (VOF-CSF) was validated with standard two-phase benchmark problems in literature [18, 19].

A common issue in the numerical simulation of two-phase flows is "spurious" or "parasitic" currents that is the generation of non-physical flow at the interface caused by an imbalance between the discrete surface tension force and the pressure gradient term that is originated from the inaccurate calculation of interface curvature [20-22]. The errors caused by the spurious current are significant in surface-tension-controlling flows (i.e. microfluidic systems) but generally small in inertia-controlling flows [23, 24]. In order to suppress the spurious currents, Lafaurie et al. [20] presented a Laplacian filter to smooth the interface and providing a less steep gradient of phase fraction. The Laplacian filter, which is used in this study, transforms the initially calculated phase fraction ($\alpha_p$) into a smoothed one ($\tilde{\alpha}_p$) by a linear interpolation of the volume fraction in each cell with the volume fractions of its neighbor cells:

$$\tilde{\alpha}_p = \frac{\sum_{f=1}^{n} \alpha_f S_f}{\sum_{f=1}^{n} S_f} . \tag{7}$$

where $p$ and $f$ are cell index and surface index, respectively. $\alpha_f$ is linearly-interpolated volume fraction at the center of the cell face, and $S_f$ is cell surface area. The implementation of the Laplacian smoother has been validated with the benchmark problems in previous studies in literature [18, 25].

The governing equations [Eqs. (1-4)] were discretized by the finite volume method and numerically solved in OpenFOAM. In order to calculate the pressure-velocity coupling, a PIMPLE algorithm was implemented that combines the Pressure-Implicit Split-Operator (PISO) and the Semi-Implicit Method for Pressure-Linked Equations (SIMPLE) algorithm. The Courant number ($C=u\Delta t/\Delta x$) was set to a limit of 0.5 to help the convergence of the solution, where $\Delta t$ is the simulation time step and $\Delta x$ is the axial spacing of the grid of the computational model.

2.2 Computational domain and boundary conditions

Numerical simulation of the two-phase flow in a tube including a porous-tube-insert is inherently a three-dimensional (3-D) system. However, for a vertical flow, due to the neutral effect of gravity in the radial direction, it is reasonable to approximate the 3-D system to a two-dimensional (2-D) axisymmetric one. Figure 1(a) shows the simplified 2-D system using ring-shape pores with a pore width of $w'$ and a wire thickness of $\delta'$ to provide an equal flow area and capillary pressure as the corresponding 3-D system with square pores having a pore width of $w$ and a wire thickness of $\delta$. The geometric conversion is required to satisfy the criteria of equal flow area and capillary pressure [14] as below,

$$w' = 0.5w , \tag{8}$$

$$\delta' = \delta + \frac{\delta^2}{2w} . \tag{9}$$

In the 2-D system, both the pore width and wire thickness are 100 $\mu$m, which corresponds to a pore width of 200 $\mu$m and a wire thickness of 82.84 $\mu$m in an equivalent 3-D system.

Figure 1(b) shows the computational domain of the 2-D vertical upward slug flow in a vertical tube with a porous-tube-insert (PTI) with a closed frontal end. The computational domain consists of a multiscale coupled grid system as shown in Figs. 1(c-e); A quasi-uniform mesh size of 100 $\mu$m is generated in the bulk flow region along the axial direction [Fig. 1(c)] except near the tube wall [Fig. 1(d)] and pores [Fig. 1(e)]. The height of the wall boundary layer mesh was gradually refined along the radial direction toward the tube wall ranging from 100 $\mu$m in the bulk region to 5 $\mu$m in the first mesh of the boundary layer on the tube wall [Fig. 1(d)]. The refined meshes were also created near the pores where the mesh size is 16.67×16.67 $\mu$m and twenty-eight meshes are used within a square pore [Fig. 1(e)]. From a mesh dependency analysis using average heat flux, it was found that 429,400 mesh was sufficient to ensure the reasonable accuracy.

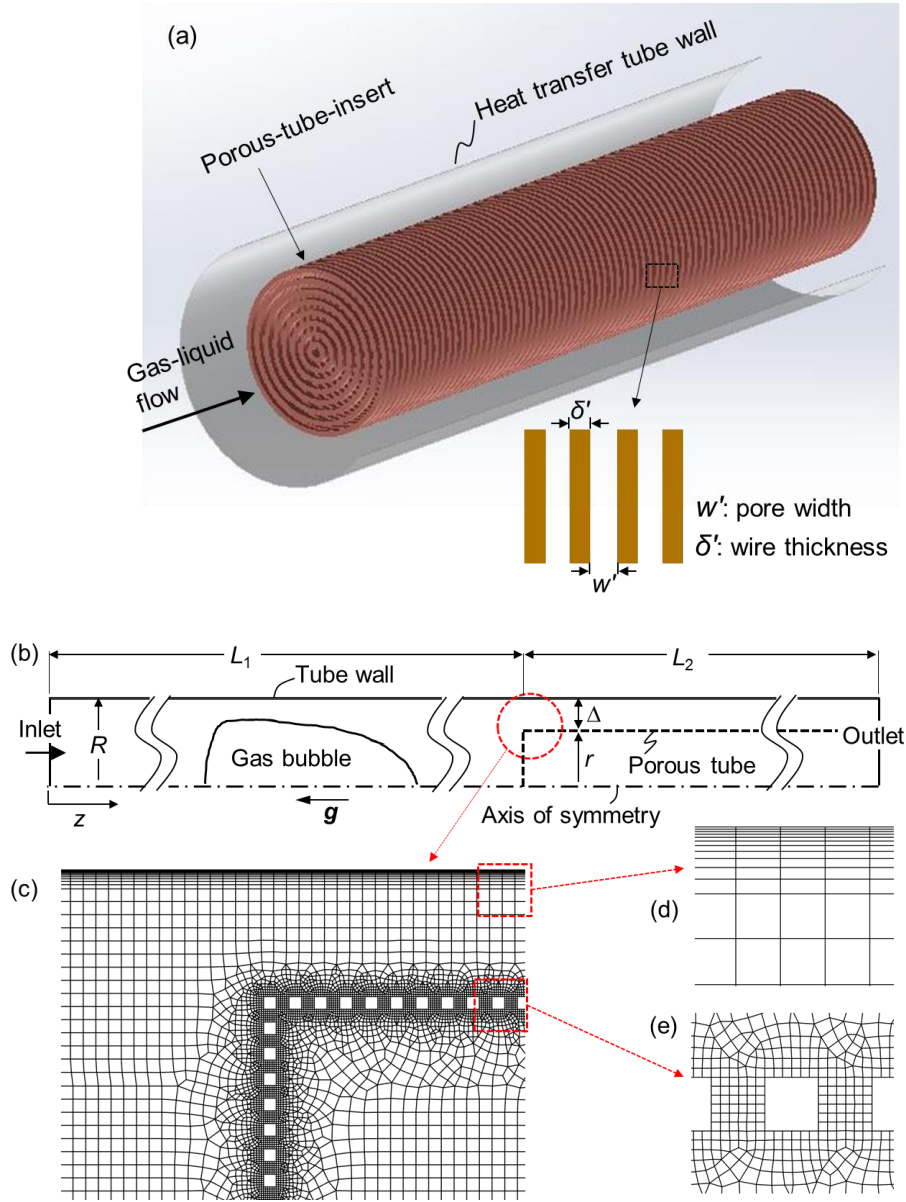

**Figure 1**. (a) 2-D axisymmetric system for vertical upflow slug flows in a tube with a porous-tube-insert with closed frontal side; and (b) computational domain of the slug flow; and (c) multiscale grid system

used in the computational domain with refined meshes (d) near the tube wall and (e) in the vicinity of the pores

Table 1 lists the thermophysical properties of the working fluids at 20°C. Table 2 lists the major parameters used for the numerical simulations. The air-water mixture flow was introduced at the tube inlet at a constant temperature ($T_i$) of 20°C. A uniform velocity ($u_i$) and a homogenous void fraction ($\alpha_i$) was used at the tube inlet, which allows small gas bubbles to grow into slug bubbles, and actual void fraction to be determined during the numerical simulation. The no-slip condition was applied on each pore surface and the tube wall. It is worth mentioning that due to the absence of actual triple-phase (gas-liquid-solid) point, the wall contact angle is not important in this study. The liquid film thicknesses around the gas bubbles were in the range of 20$\mu$m ~ 800 $\mu$m, where the van der Waal attraction between the liquid and solid phases is negligible for such a thick liquid film [8, 14]. The atmospheric pressure was set at the tube outlet. This study used a constant wall temperature of 10°C, which was 10°C colder than the inlet fluid's temperature, so that the air-water mixture flow is cooled by the tube wall.

**Table 1**. Thermophysical properties of working fluids at 20°C used for the numerical simulations

| Fluid | Density, $\rho$ [kg/m$^3$] | Kinematic viscosity, $v$ [m$^2$/s] | Specific heat, $c_p$ [J/kgK] | Thermal conductivity, $k$ [W/mK] | Surface tension, $\sigma$ N/m |
|---|---|---|---|---|---|
| Water | 998 | 10$^{-6}$ | 4182 | 0.600 | 0.072 |
| Air | 1.2 | 1.480×10$^{-5}$ | 1006 | 0.024 | - |

**Table 2.** Major parameters used for the numerical simulations

| Parameter | Value |
|---|---|
| Tube length, $L$ [mm] | 900 |
| Length of bare tube section, $L_1$ [mm] | 500 |
| Length of porous-insert-tube section, $L_2$ [mm] | 400 |
| Tube radius, $R$ [mm] | 3.150 |
| Porous tube radius, $r$ [mm] | 1.650 |
| Wire thickness, $\delta'$ [mm] | 0.100 |
| Pore width, $w'$ [mm] | 0.100 |
| Tube wall temperature, $T_w$ [°C] | 10 |
| Inlet temperature, $T_i$ [°C] | 20 |
| Inlet void fraction of the homogenous two-phase flows, $\varepsilon_i$ [-] | 0.400 |
| Inlet velocity of the homogenous two-phase flows, $u_i$ [m/s] | 0.068 |

In order to simulate the complex hydrodynamics of the two-phase flows, a transient simulation with a time step of $10^{-5}$ s was used. Gravitational field was also applied opposite to the flow direction to create an upward flow configuration.

## 3. Results and discussion

Five different cases were considered in this study as shown in Fig. 2. A liquid-only (single-phase) flow was selected as a baseline (Case 1), and a two-phase slug flow in a bare tube was considered as another baseline for two-phase slug flows (Case 2). Three cases using either a porous (Cases 3, 5) or solid (Case 4) insert with the same diameter and length were used to study the hydrodynamic characteristics and heat transfer enhancement mechanisms:

- Case 3: A porous-tube-insert (PTI) with a closed front suspended in a tube is used to create elongated ring-shaped slug bubbles in the annular gap between the PTI and tube wall.
- Case 4: A solid-insert (SI) suspended in a tube is used to create elongated ring-shaped slug bubbles in the annular gap between the SI and tube wall. Note that the liquid flow can't freely flow through the SI for Case 4 but can freely flow through the PTI for Case 3.
- Case 5: A porous-tube-insert (PTI) with an open front suspended in a tube is used to capture the slug bubbles stay inside the porous tube.

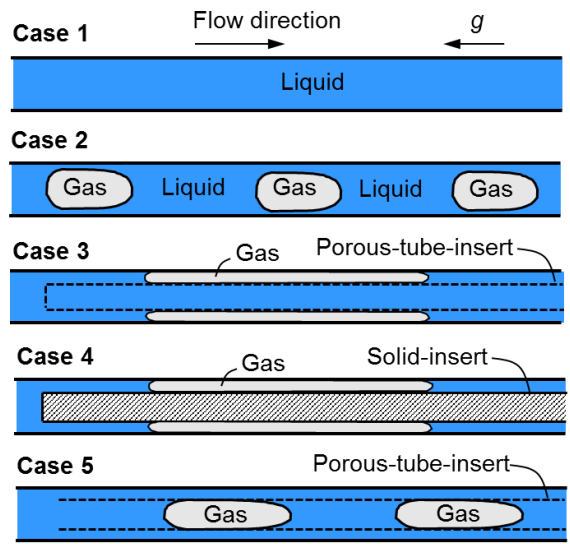

**Figure 2**. Five different cases used for comparison of the convective heat transfer performance

3.1 Two-phase slug flow in bare tube (Case 2)

Two-dimensional axisymmetric simulations of two-phase flows of water and air in a tube with a diameter of 6.3 mm and a length of 900 mm (~143$D$) were carried out. A mixture velocity ($u_i$) of 0.068 m/s and a void fraction ($\varepsilon_i$) of 0.4 at the inlet were used to create a slug flow in the tube (Case 2 in Fig. 2). The calculated liquid and gas Reynolds numbers are 257 and 12, respectively and therefore the two-phase flow is laminar. The inlet temperature of the flow of 20 °C and a constant tube wall temperature of 10 °C were used to cool the fluid by the tube. The baseline case for a single-phase (liquid-only) flow (Case 1 in Fig. 2) with the same conditions was also simulated for comparison.

Figure 3(a) shows the streamlines of the liquid flow around a gas bubble in a fully-developed flow region. The streamlines of the liquid flow were drawn in a fixed frame of reference (FFR). It is apparent that the falling liquid film through the gap between the tube wall and bubble creates a vortex behind the bubble tail. The temperature contour [Fig. 3(b)] indicates that the temperature inside the bubble is relatively lower than that of the liquid slugs. Because of the vortex pushing the hot liquid toward the wall, the temperature in the liquid slug behind the bubble tail is high and uniform. Similar qualitative temperature distributions were reported in the previous studies in literature for slug (Taylor) flow in microchannels [6, 7, 26].

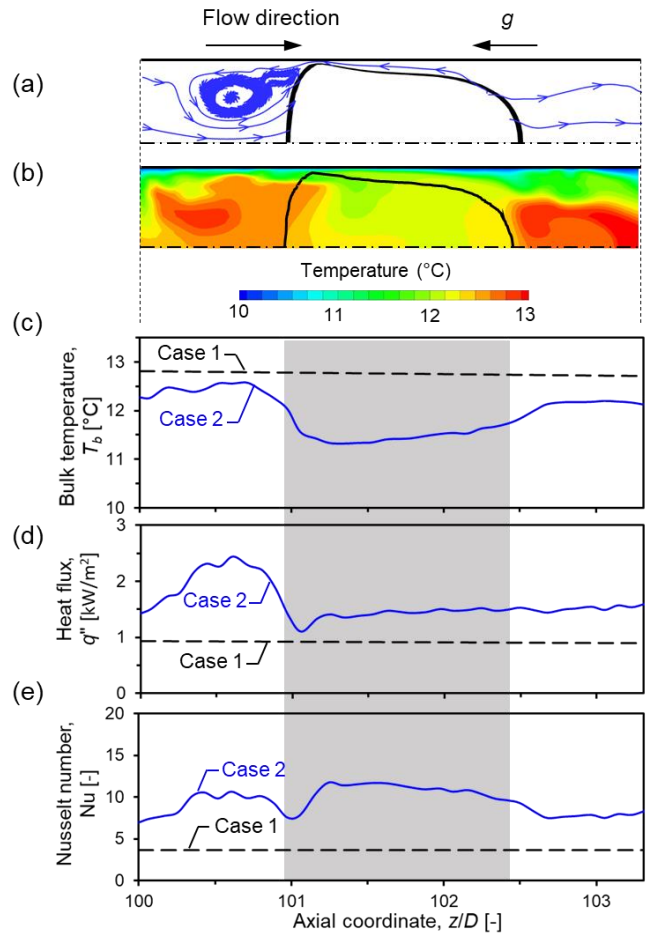

**Figure 3**. (a) Streamlines of liquid flow. (b) Temperature contour around a gas bubble. Axial variations of (c) the bulk temperature, (d) local wall heat flux, and (e) local Nusselt number in the fully-developed flow region along the axial distance ($z$) from the inlet

Figure 3(c) shows the bulk temperature ($T_b$) of the two-phase flow in the fully-developed region. The bulk (mean) temperature at a certain axial location is defined as

$$T_b = \frac{\int_0^R \rho |u| c_p T (2\pi r) dr}{\int_0^R \rho |u| c_p (2\pi r) dr}, \quad (10)$$

where $R$ is the tube radius and $r$ is the radial coordinate. In order to properly account the contributions of the falling film and vortex region, the absolute value of the axial velocity ($u$) was used for the bulk temperature calculation [27]. It is worthwhile to mention that that the radial velocity is ignored in Eq. (10), because the magnitude of the axial velocity dominates the radial component. Figure 3(c) shows a higher bulk temperature in the liquid regions than that of the bubble (shared area) as seen in Fig. 3(b) and the bulk temperature of the baseline liquid-only flow (Case 1) is higher than that of the two-phase flow because of the inferior heat transfer of the liquid-only flow upstream of the region shown in the figure. Details of the results of the heat transfer are discussed below.

Figure 3(d) shows the variation of the local wall heat flux ($q''$). The wall heat flux was calculated by the conduction through the first liquid layer (5 $\mu$m thick) fixed on the tube wall as below

$$q'' = -k \left.\frac{\partial T}{\partial r}\right|_{r=R}, \quad (11)$$

where $r$ is the radial coordinate and $k$ is the liquid thermal conductivity. It is seen from Fig. 3(d) that the heat flux in the liquid behind the bubble tail is relatively high due to the strong mixing by the vortex. The heat flux decreases toward the bubble region but is still higher than that of the liquid-only flow.

The local Nusselt number (Nu) shown in Fig. 3(e) is calculated by

$$\mathrm{Nu} = \frac{q'' D}{k_L (T_b - T_w)}. \quad (12)$$

The Nusselt number is higher in the entire region than that of the liquid-only flow (Nu=3.66 for Case 1). The strong mixing in the vortex region and the thin liquid film in the bubble region increases the local Nusselt number. The heat flux in the bubble region is, however, relatively low because of the lower bulk temperature [Fig. 3(c)]. The high bulk temperature and convective heat transfer coefficient in the vortex region greatly enhances the convection heat transfer.

For a quantitative comparison of the heat transfer, the average Nusselt number was calculated using the averaged heat flux and bulk temperature over the entire slug unit consisting of a bubble and a liquid slug as below

$$\mathrm{Nu}_{ave} = \frac{q''_{ave}}{(T_{b,ave} - T_w)} \frac{D}{k_L}, \tag{13}$$

$$q''_{ave} = \frac{1}{L_u} \int_0^{L_u} q'' \, dz, \tag{14}$$

$$T_{b,ave} = \frac{1}{L_u} \int_0^{L_u} T_b \, dz, \tag{15}$$

where $L_u$ is the total length of the entire slug unit. The averaged Nusselt number of the slug flow was 9.37 that is 156% higher than that of the equivalent liquid-only flow.

The current CFD model was validated for void fraction, bubble velocity, and flow regimes of adiabatic two-phase (gas-liquid) flows in the authors' previous works [28-30]. For the heat transfer of two-phase slug flows in vertical tubes, Dong and Hibiki [31] identified more than 200 experimental data through a comprehensive literature survey. Using the experimental data bank and the theoretical Reynolds and Chilton–Colburn analogies, they presented a correlation to predict the heat transfer coefficient (or Nusselt number) for the slug flows in vertical tubes, which successfully predicted 95.1% of the data within ± 30% error. The correlation for laminar flow ($Re_L$<2300) is given by

$$\frac{Nu_{2\varphi}}{Nu_{1\varphi}} = (1-\varepsilon)^{0.339}\left(1+\frac{4.65}{X^{0.409}}\right), \tag{16}$$

$$Nu_{1\varphi} = 1.86\left(Re_L \, Pr_L \, \frac{D}{L}\right)^{1/3}\left(\frac{\mu_b}{\mu_w}\right)^{0.14}, \tag{17}$$

where $Nu_{2\varphi}$ and $Nu_{1\varphi}$ are the two-phase and liquid-phase Nusselt numbers, respectively; $\mu_w$ and $\mu_b$ are the liquid viscosities calculated based on the temperatures of the tube wall and bulk fluid, respectively. $\varepsilon$ is the void fraction of a two-phase flow and $X$ is the Martinelli parameter and is defined as

$$X = \frac{\dot{m}_L}{\dot{m}_G}\sqrt{\frac{\rho_G}{\rho_L}} = \left(\frac{1-x}{x}\right)^{0.9}\left(\frac{\rho_G}{\rho_L}\right)^{0.5}\left(\frac{\mu_L}{\mu_G}\right)^{0.1}, \tag{18}$$

where $x$ is the gas (vapor) quality; and $\dot{m}_L$ and $\dot{m}_G$ are the mass flow rates of liquid and gas phases, respectively.

Fifteen different laminar air-water slug flow cases with various inlet conditions (inlet mixture velocity and homogenous void fraction) were numerically simulated for model validation purpose. The average Nusselt numbers were calculated using Eq. (13) in the numerical simulation and compared with the values predicted by the Dong and Hibiki's correlation [Eq. (16)] within an error band of ±15% as shown in Fig. 4.

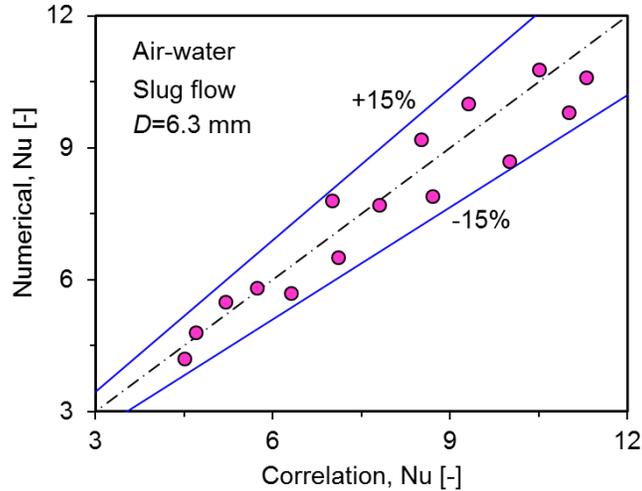

**Figure 4**. Comparison of the computed average Nusselt number of two-phase slug flows with those predicted by Dong and Hibiki's correlation [31]

3.2 Convective heat transfer enhancement of slug flows using suspended tube inserts (Cases 3-5)

3.2.1 Hydrodynamic characteristics

Figure 5 shows the slug bubble flow in a bare tube [Fig. 5(a)] and the modulated flows using tube inserts [Fig. 5(b-d)]. Because of the axisymmetric nature of the vertical flows, only the computational domains shown in Fig. 5 were used for the simulation. It is noteworthy that the bubble could not penetrate through the small pores (200 μm) of the porous tube (Cases 3 and 5) for the conditions used for this study, resulting in a capillary-assisted phase separation. Therefore, in Case 3, the gas bubble is forced to flow through the narrow annular gap, elongating the bubble in a ring shape. The same shape change of the bubble occurs in Case 4 using a solid insert [Fig. 5(c)]. To deal with all possible designs using a tube insert, Case 5 using a porous tube with an open front was considered to keep the bubble inside the porous tube instead of the annular gap (Case 3).

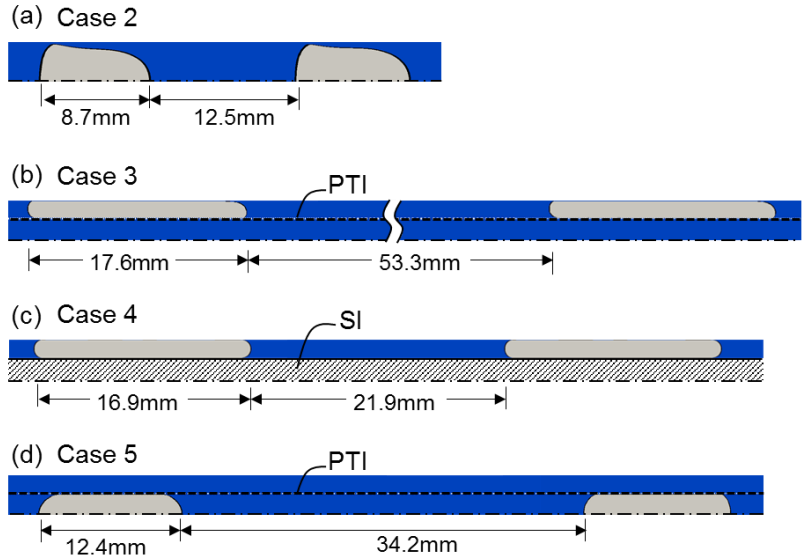

**Figure 5**. Four different designs considered for computational simulation: (a) bare tube (Case 2); (b) porous-tube-insert with a closed front (Case 3); (c) solid-insert (Case 4); and (b) porous-tube-insert with an open front (Case 5)

      The elongated bubble in Cases 3 and 4 is almost twice longer than that of Case 2 because of the bubble confinement in the annular gap and the bubble in Case 5 is about 1.4 times of that of Case 2. The liquid slug is also extended for all the cases. The liquid slug in Case 3 is maximum (53.3mm) and is more than four times of that in Case 2. It is also true for the liquid slugs in Cases 3-5 which are much longer than that of Case 2. The elongation of the liquid slugs in the cases using the PTI (Cases 3 and 5) are due to the liquid circulation through the porous-insert as discussed below.

      Figure 6 shows the streamlines of the liquid flow around the bubble for Case 3 in a fixed frame of reference. The large difference in the densities of the liquid and gas phases and the resulting buoyancy force create a counterflow of the bubble moving upward in the annular gap and liquid moving downward inside the porous tube as shown in Fig. 6(a), resulting in an internal circulation of the liquid flow through the porous insert.

      Figures 6(b-d) show the magnified views of the flow patterns at the different locations around the bubble. In Fig. 6(c) (region 2), the bubble stays outside the porous tube because the gas pressure in the bubble is too low to break the gas-vapor interface (meniscus) formed in the micropores of the porous tube, while the liquid can freely flow through the pores. The liquid film thickness is about 17 $\mu$m between the pore surface and the bubble.

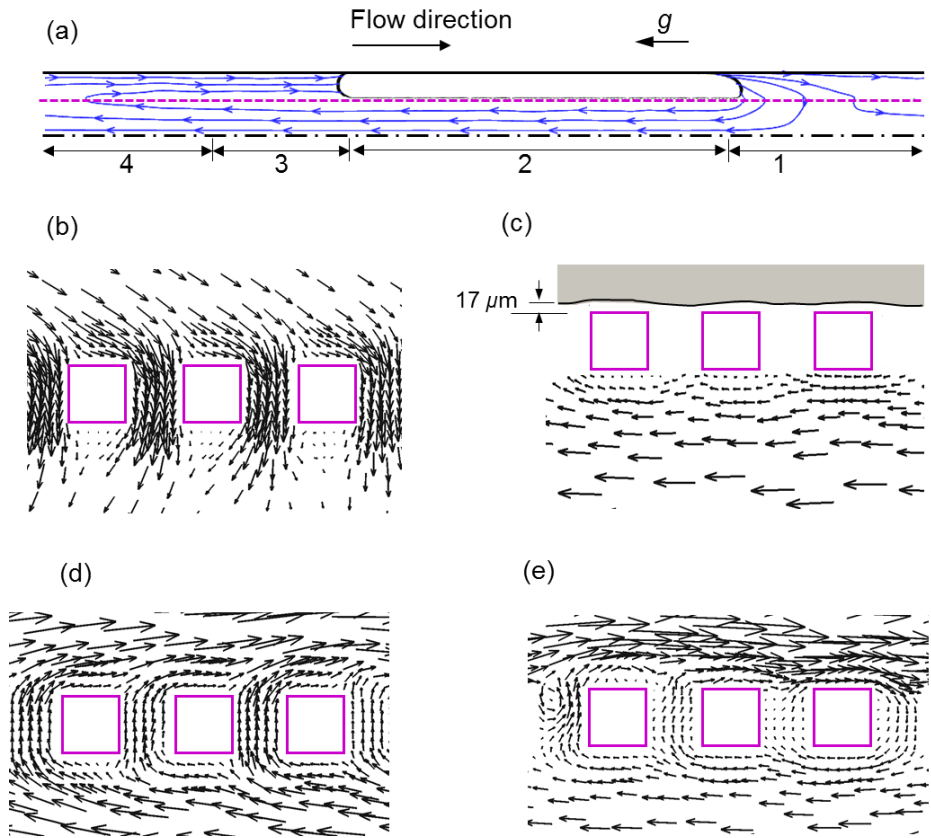

**Figure 6**. (a) Streamlines of the liquid flow around a gas bubble outside the porous-tube-insert with a closed front (Case 3); and enlarged flow vectors near the pores of the porous tube in (b) region 1; (c) region 2; (d) region 3; and (e) region 4

Figure 7 shows the streamlines of the liquid flow around the bubble for Case 5 in a fixed frame of reference. Case 5 has the bubble flowing inside the porous tube, which was introduced through the open front of the porous tube. Similar to Case 3, Case 5 also has a liquid circulation but through the annular gap.

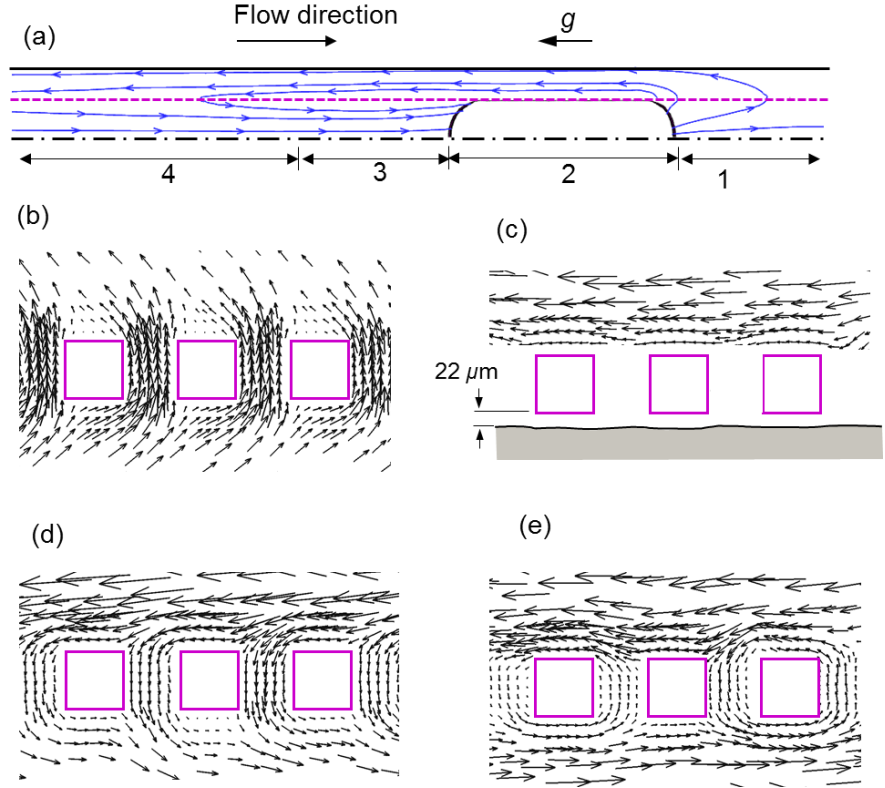

**Figure 7**. (a) Streamlines of the liquid flow around a gas bubble inside the porous-tube-insert with a open front (Case 5); and enlarged flow vectors near the pores of the porous tube in (b) region 1; (c) region 2; (d) region 3; and (e) region 4

To study the effects of the inserts on the flow modulation, two locations [A and B in Fig. 8(a)] were selected for comparison of the cross-sectional velocity profile in the middle of the bubble and liquid slug for each case, respectively.

Figures 8(b-c) compare the cross-sectional axial velocity profiles for different cases in the middle of the bubble. As shown in Fig. 8(b), the bubble travels faster in the cases with PTI (Cases 3 and 5) than the other cases (Cases 1, 2 and 4). This is because of the fast liquid flowing backward through either the porous tube (Case 3) or the annular gap (Case 5), which results in the fast bubble moving forward. In contrast, in Case 4 with a solid insert, the bubble and liquid slug flow slowly in the confined annular gap without a substantial liquid flowing backward. The similar results are also observed in the bare tube (Case 2).

In Fig. 8(c), the cross-sectional axial velocity profiles in the liquid slug for different cases are compared. It is obvious that the velocity profiles of Cases 3 and 5 using a PTI, because of the liquid flow and phase separation, are dramatically different from the typical parabolic profile of a developed flow in a bare tube (Case 1 and 2). The maximum velocities in Cases 3 and 5 with a PTI are now shifted toward the tube wall from the tube center of the bare tube (Case 1 and 2), and the velocity profiles have both positive (forward) and negative (downward) velocity regions unlike Cases 1, 2 and 4.

In Case 4 with a solid insert, although the maximum velocity is increased due to the reduced flow area compared to that of the bare tube, it is much less than those in the PTI cases. This attributes to the existence of significant downward liquid flow in the PTI cases as explained above. Note that the velocities at the porous tube wall are zero because of the no-slip boundary condition.

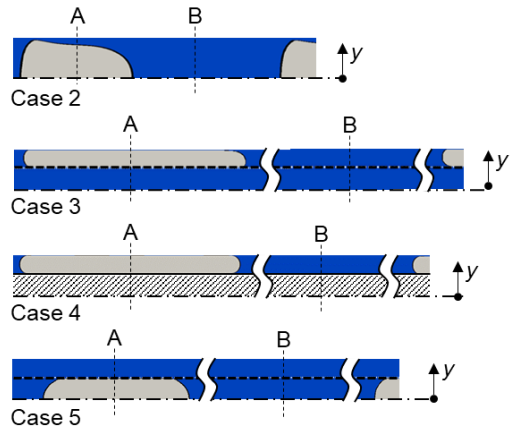

(a)

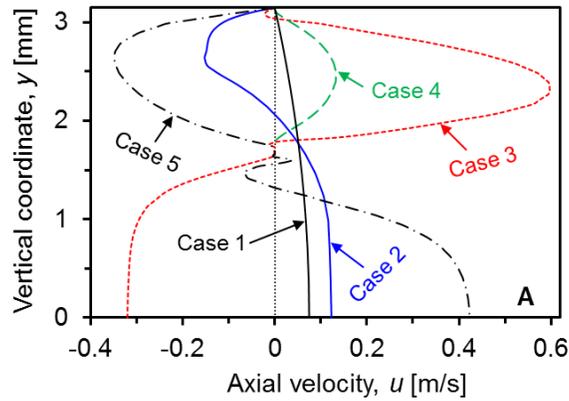

(b)

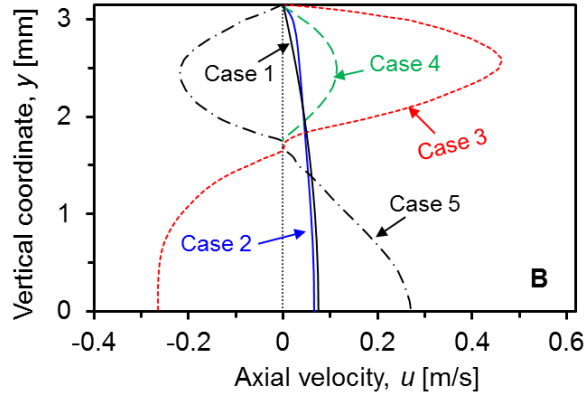

(c)

**Figure 8.** (a) Selected tube cross-sections in different cases. The axial velocity profiles at (a) the cross section "A" in the bubble; and (b) the cross section "B" in the liquid slug

  The convective heat transfer is greatly affected to the fluid's velocity gradient on the wall as stated by the Reynolds analogy [32]. Figure 9 shows the comparison of the averaged axial velocities of the liquid slugs between two consecutive bubbles, specifically in the annular gaps for Cases 3-5. The average velocity of the fully developed single phase flow (Case 1) is also shown as the baseline. Note that the negative values indicates a downward flow for Case 5. It is obvious that the liquid velocity for Case 4 with a solid insert is higher than that of Case 2 using a bare tube due to the reduced flow area.

  For Cases 3 and 5 using a PTI, however, the matter is more complicated. In Case 3, the axial velocity decreases in a short region close to the nose of the following bubble because of the radial flow leaking from annular gap to the porous tube [Fig. 6(b)] but gradually increases due to the liquid infiltration from the porous tube [Fig. 6(c-d)]. In Case 5, the opposite results to Case 3 are observed and therefore the liquid flows downward in the entire annular gap. Although the flow directions in the annular gap for Cases 3 and 5 are opposite each other, the velocities are higher than that in the bare tube (Case 2), especially behind the bubble tail.

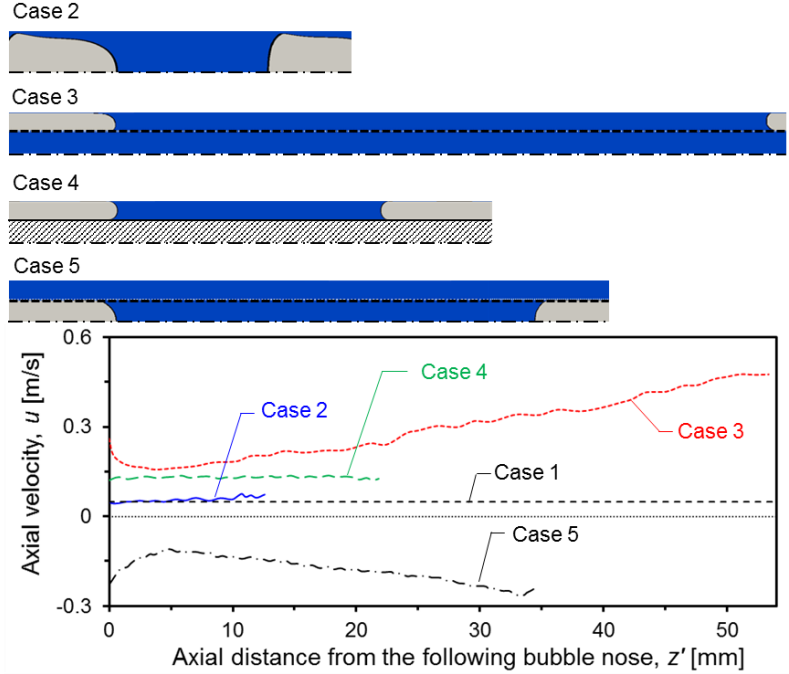

**Figure 9.** Axial velocity variation in the liquid slug between two consecutive bubbles for the different cases

Figure 10(a) shows the trace of the gas-liquid interface of a fully-developed bubble in the bare tube (Case 2). The interface of the bubble is defined by a line with a phase fraction ($\alpha$) of 0.5. The local liquid thickness ($\delta$) between the interface and tube wall is illustrated in Fig. 10(a). The local liquid film thicknesses along the axial distance ($z'$) [Fig. 10(a)] from the bubble tail are shown in Fig. 10(b). Note that the film thicknesses are shown in a logarithmic scale. It is seen that the confined bubbles in Cases 3 and 4 have longer length and much thinner film thickness than those in the bare tube (Case 2). For Case 5, the bubble is slightly elongated as compared to that in Case 2 and the film thickness equals to the thickness of the annular gap.

Figure 10(c) shows the area-averaged liquid film thicknesses for different cases, which is calculated by

$$\delta_{ave} = \frac{1}{L_b} \int_0^{L_b} \delta \, dz'. \tag{19}$$

It is seen that the average film thickness is significantly reduced in Cases 3 and 4 where the film thicknesses are about 1/10 and 1/20 of that of the bare tube, respectively. Note that the average liquid film thickness of Case 5 is much larger than that of Case 2, which provides more liquid flow of high temperature required for the convective heat transfer.

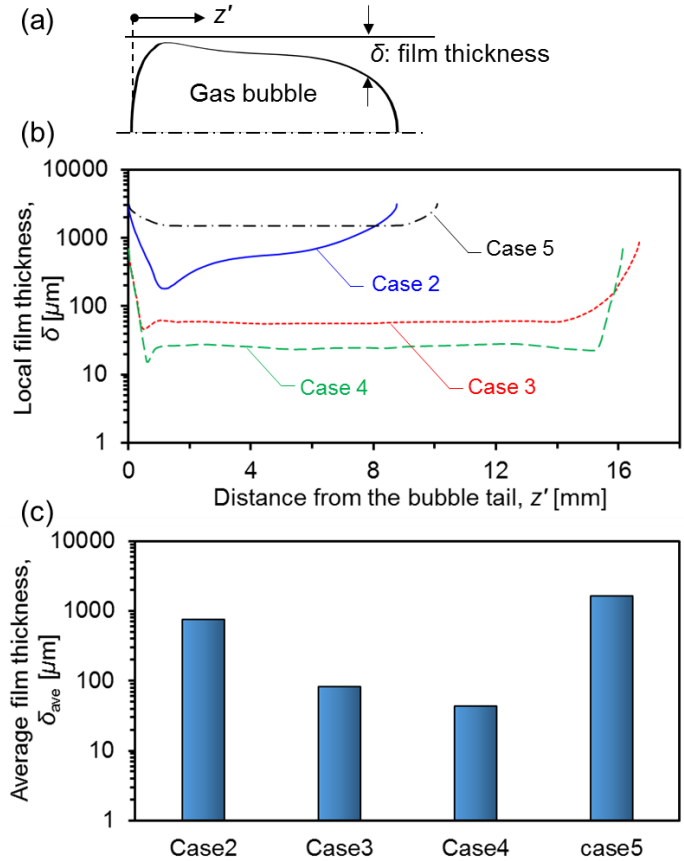

**Figure 10.** (a) A gas bubble in a fully-developed slug flow in the bare tube (Case 2); and variation of the (b) local and (c) area-averaged film thickness along a gas bubble in the different cases

### 3.2.2 Thermal characteristics

The temperature contours in the gas bubble and the adjacent liquid slugs for different cases are shown in Fig. 11. In Cases 3 and 4, the temperatures of the bubble and liquid film below the bubble quickly equilibrate to the wall temperature because of the small thermal capacity (low density of the bubble and small volume of the thin liquid film), while the temperatures of the adjacent liquid slugs near the wall are higher than the wall temperature because of the large thermal capacity.

The bubble-induced liquid circulation in Cases 3 and 5 is a very effective way of mixing hot and cold fluids. In Case 5 with a desirable flow arrangement for the convective heat transfer enhancement, the bubble pushes the hot liquid in the porous tube out to the annular gap adjacent to the tube wall. Then, the hot liquid flows downward, while being cooled by the tube wall, and back into the porous tube behind the bubble.

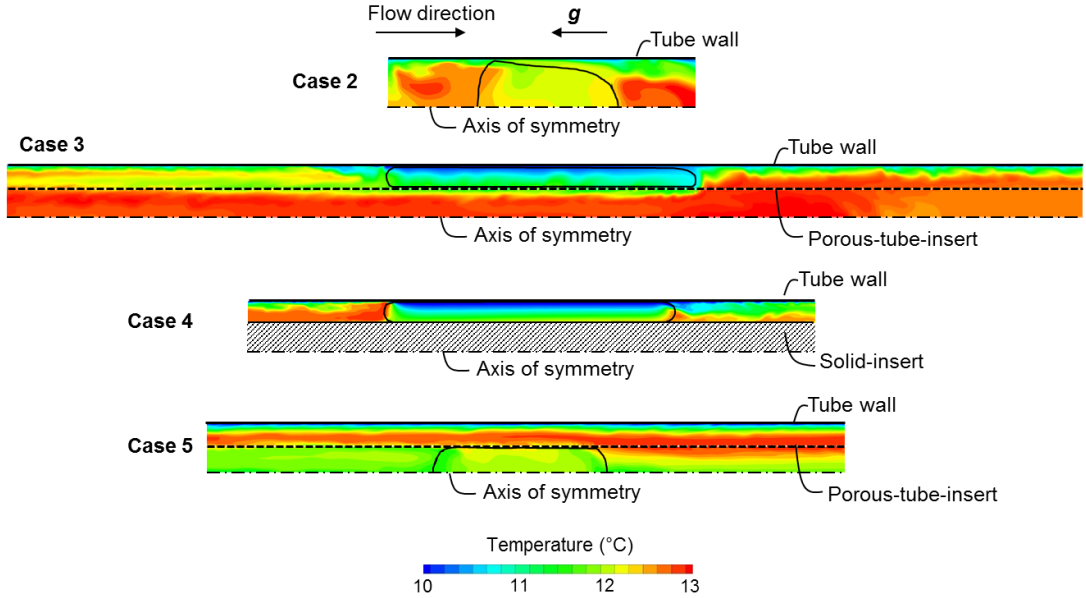

**Figure 11.** Temperature contours in a gas bubble and the adjacent liquid slugs in different insert cases

The variations of the bulk temperatures in the slug units for all cases are shown in Fig. 12(a). The bulk temperatures are calculated using Eq. (10) in the fully-developed region ($98 < z/D < 111$). In Fig. 12(a), the big drops of the bulk temperatures in the bubble region are observed for all cases except Case 5. Detailed discussion on the results for Cases 1 and 2 was given in Section 3.1.

Figure 12(b) shows the local wall heat fluxes in the slug units for all cases, which were calculated from Eq. (11). It is observed that the heat fluxes reach the peak behind the bubble tail for all cases except Case 5, because of the highest liquid velocity near the wall (Fig. 9) and then rapidly drop in the bubble region lined with a thin liquid film, to a lower value than that of the liquid-only flow for Cases 3 and 4. In contrast, in Case 5, the heat flux is the highest near the bubble head where the liquid velocity near the wall is increased and quickly drops into the liquid slug of the low velocity (Fig. 9). It is noteworthy that for the cases using the PTI, small fluctuations of the wall heat fluxes in the liquid slug regions due to the small jet flows through the pores of the PTI (Figs. 6 and 7).

Figure 12(c) presents the local Nusselt numbers in a slug unit [Eq. (12)] which was calculated in using Eq. (12). Note that for Case 4 using a solid insert, the hydraulic diameter, which is two times of the annular gap, was used as the length scale for the Nusselt number. In general, the trends of the variations of the Nusselt number are similar to those of the heat flux variations.

The average heat fluxes and Nusselt numbers over the slug unit for all cases are normalized with Case 1 and compared in Fig. 12(d). It is clearly shown that Cases 3 and 5 with the PTI stand out for both heat flux and Nusselt number. The average Nusselt number for Cases 3 and 5 are 17.23 and 17.91, respectively, which are about 5 times greater than that of the liquid-only flow (Nu=3.66). Case 4 using a solid insert is the worst insert design (Nu=3.74) with no major gain as compared to the liquid-only flow. The liquid circulation through the PTI increases the velocity near the wall, which leads to the convective heat transfer enhancement.

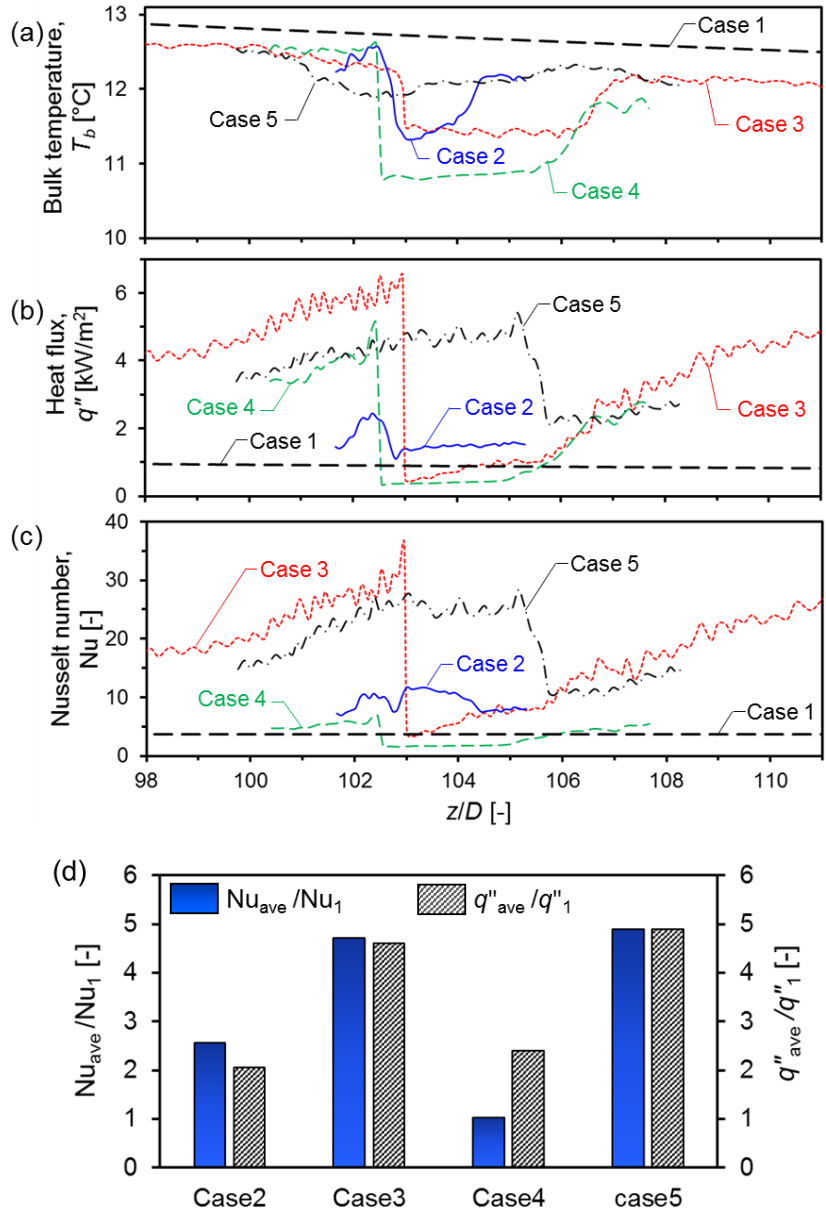

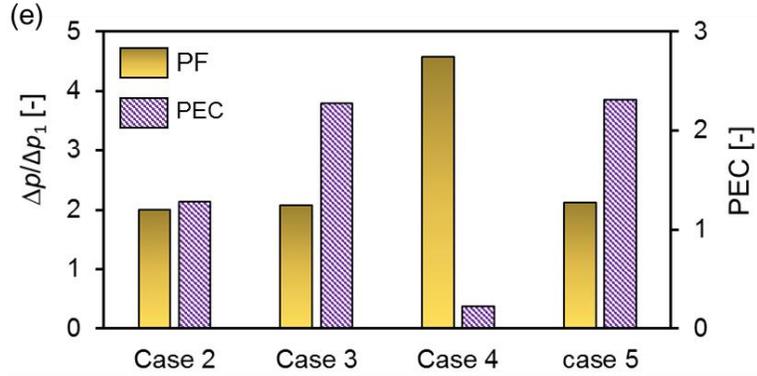

**Figure 12**. Variations of (a) the local bulk mean temperature; (b) the local heat flux; (c) the local Nusselt number. Comparison of (d) the normalized average Nusselt number and heat flux; and (e) the normalized pressure drop and performance evaluation criteria (PEC) calculated over a slug unit

Figure 12 (e) compares the computed pressure drop for all cases that were normalized with Case 1 (pure liquid flow) as well as the performance evaluation criteria (PEC) to consider both heat transfer enhancement and pressure drop penalty:

$$\text{PEC} = \frac{(\text{Nu}/\text{Nu}_1)}{(\Delta p/\Delta p_1)}, \tag{20}$$

It was found that the overall performances of the porous-tube-inserts (Cases 3 and 5) are much higher than the other cases (Cases 2 and 4). In particular, using the solid-insert (Case 4) led to an excessive pressure drop of about 4.5 times higher than that of the pure liquid flow, resulting in a poor PEC of about 0.2. On the other hand, the maximum PEC of 2.31 was achieved for Case 5.

The annular gap dimension (or porous tube diameter) for Cases 3 and 5 using the PTI is an important design parameter in determining the length of bubble and liquid slug and affecting the heat transfer performance. In the following section, numerical simulations were performed for different annular gap dimensions in a fixed heat transfer tube diameter to find an optimum gap dimension for the maximum overall heat transfer.

3.3　Effect of gap dimension on slug flows and heat transfer

Table 3 lists the five different annular gap dimensions ($\Delta$) in the PTI cases used for the numerical simulations. In the table, $R$ and $r$ are the radii of the heat transfer and porous insert tube, respectively. The annular gap ($\Delta$) ranges from 0.6 to 2.4 mm with a fixed heat transfer tube diameter of 6.3mm. An aspect ratio (AR) is the ratio of the bubble flow area (BFA) to the tube diameter:

$$\text{AR} = \frac{\text{BFA}}{2R}. \tag{21}$$

The bubble flow (BFA) in Case 3 is two times of the annular gap (2Δ), whereas the BFA in Case 5 is the porous tube diameter (=2r). In Table 3, the aspect ratio (AR) considered for both Cases 3 and 5 has almost the same ranges of 0.2< AR <0.77.

**Table 3**. Values of the annular gap, bubble flow area, and aspect ratio of Cases 3 and 5 used for a parametric study

| Case# | $R$ (mm) | Δ (mm) | $r$ (mm) | AR (-) |
|---|---|---|---|---|
| 3a | 3.15 | 0.6 | 2.45 | 0.19 |
| 3b | 3.15 | 1.0 | 2.05 | 0.32 |
| 3c | 3.15 | 1.4 | 1.65 | 0.44 |
| 3d | 3.15 | 1.8 | 1.25 | 0.57 |
| 3e | 3.15 | 2.4 | 0.65 | 0.76 |
| 5a | 3.15 | 0.6 | 2.45 | 0.77 |
| 5b | 3.15 | 1.0 | 2.05 | 0.65 |
| 5c | 3.15 | 1.4 | 1.65 | 0.52 |
| 5d | 3.15 | 1.8 | 1.25 | 0.40 |
| 5e | 3.15 | 2.4 | 0.65 | 0.21 |

As discussed in Section 3.2.2, the convective heat transfer occurs mainly in the liquid slug regions. Figure 13(a-b) illustrates the dimensional changes of the fully-developed bubble and liquid slug for different the aspect ratio (AR) of Cases 3 and 5. It is observed in Figs. 13(a-b) that by increasing the aspect ratio (AR), the bubble and liquid slug are shortened for both cases. The lengths of the liquid slug and bubble for different aspect ratio (AR) are presented in Fig. 13(c). The lengths of the liquid slugs and bubble decreases with increasing aspect ratio and approaches to that of the bare tube (Case 2). More elongated bubble (for smaller aspect ratio) creates a stronger liquid circulation in a longer liquid slug. Note that in Case 5e, a small inclined baffle was used at the entrance of the porous tube to guide the bubbles into the porous tube.

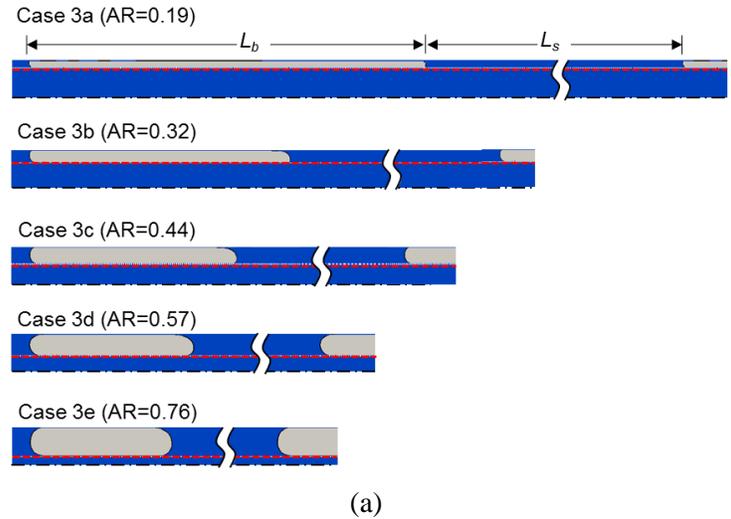

(a)

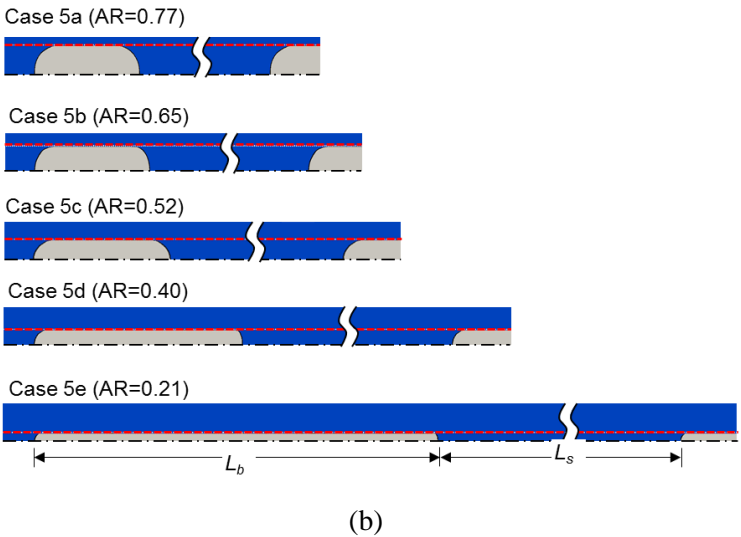

(b)

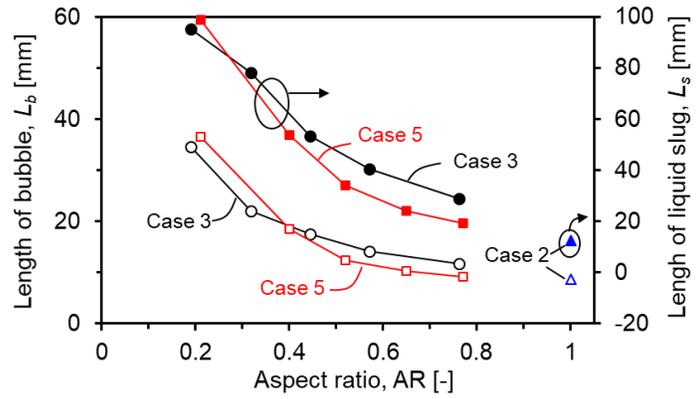

(c)

**Figure 13**. Effect of aspect ratio (AR) on flow pattern in (a) Case 3; (b) Case 5; and (c) the calculated length of gas bubble and liquid slug for different cases

As discussed earlier in Section 3.2.1, a thinner liquid film is preferred because of a small thermal resistance for the convective heat transfer but provides less thermal capacity. Figure 14 shows the computed average liquid film thickness ($\delta$) around a bubble for different aspect ratios (AR). The film thicknesses were obtained by averaging ten bubbles over an entire length of each bubble as previously discussed in Section 3.2.4. It is seen that in Case 3 as the aspect ratio (AR) increases, the film thickness also increases. The liquid films in Case 3 ranges from $30\mu$m to $95\mu$m. Note that the film thickness in Case 5 decreases, as the diameter of the porous tube increases, and therefore the annular gap decreases, according to the definition of the aspect ratio in Eq. (21).

The bubble velocity can be used to characterize the flow fields and liquid circulations in a liquid slug unit for convective heat transfer. Larger bubble velocity intensifies the liquid circulation to augment the convective heat transfer. Figure 14 shows the effect of the aspect ratio (AR) on the bubble velocity for different cases. As shown, by increasing AR, the bubble moves slower due to the increased flow area for both cases, and approach to that of the bare tube (Case 2).

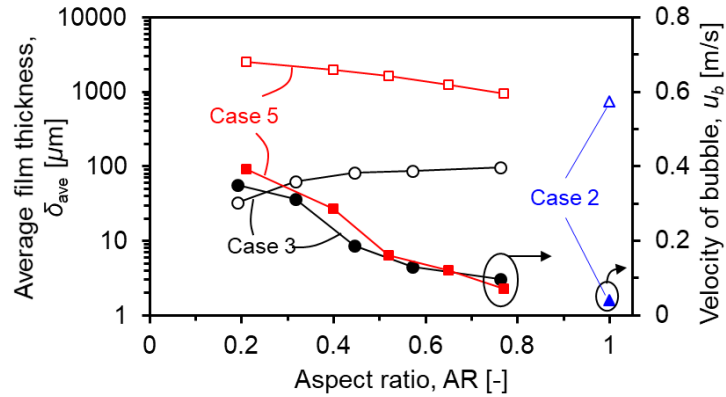

**Figure 14**. Effect of aspect ratio on the average liquid film thickness and bubble velocity

The convective heat transfer enhancement by the bubble-induced liquid circulation using a PTI can be quantified by the averaged Nusselt number. Figure 15 shows the effect of aspect ratio on the average Nusselt number for Cases 3 and 5. The averaged Nusselt numbers were calculated over a fully-developed region ($90<z/D<130$) using Eq. (13). It is seen from Fig. 15 that all cases using a PTI have higher average Nusselt number than those of liquid-only flow (Case 1) and two-phase slug flow (Case 2) in a bare tube. Interestingly, there exists an optimum aspect ratio (AR) for a maximum of the average Nusselt number in both cases. For Case3, the maximum average Nusselt number of 17.6 was achieved with AR=0.57 of Case 3d (Table 3). For Case 5, the maximum average Nusselt number of 18.24 was obtained for AR=0.65, of Case 5b.

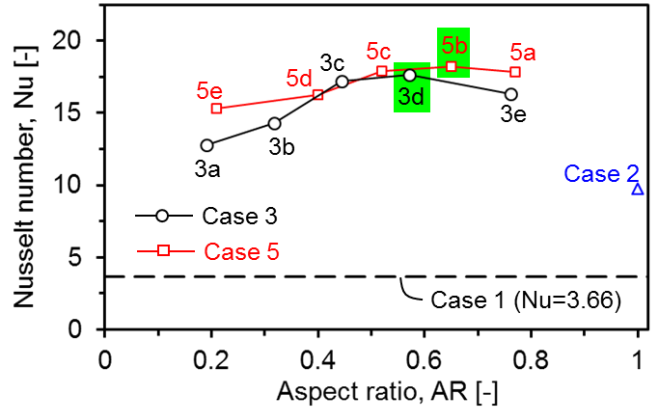

**Figure 15**. Effect of the aspect ratio on the average Nusselt number for Cases 3 and 5

## 4. Conclusion

Computational fluid dynamics (CFD) simulations were carried out to investigate the convective heat transfer enhancement in non-condensing slug flows using capillary-assisted phase separation. As such, the air-water slug flow and heat transfer without phase change in a vertical upward direction was considered where a porous-tube-insert was suspended in the main tube. Simulation results show that the capillary phase separation prevented the bubbles to penetrate through the pores of the porous tube. Therefore, depending on the front end of the porous tube, two different scenarios may occur; a closed front prevented the bubbles from entering the porous tube keeping the bubbles in the annular gap; an open front allows the bubbles to flow through the porous tube. It was found that a buoyancy-induced counter flow of the bulk liquid inside and outside of the porous tube creates internal liquid circulations-micron-scale liquid circulations through the pores and the bulk mili-scale circulations around the bubbles-leading to an increased velocity near the wall and accordingly an enhanced convective heat transfer. Finally, it was found that there exists an optimum dimension of the annular gap between the tube wall and porous tube for the maximum Nusselt number. The intriguing variation of the Nusselt number with the gap dimension is explained by the following factors:

- The magnitude of the liquid velocities near the tube wall,
- The length (area) occupied by the high liquid velocity zone near the wall (i.e. near the bubble tail region), and
- The film thickness in the bubble region. The present study shows that the average Nusselt number by using a porous-tube-insert could be about five times higher than that of a pure liquid flow in a bare tube.

## Acknowledgement

The authors extend their thanks for the financial support of the National Science Foundation (CAREER Award 1464504).